# DESIGN AND IMPLEMENTATION OF BIT TRANSITION COUNTER

Amandeep Singh[1], Balwinder Singh[2]

[1-2] Acadmic and Consultancy Services Division, Centre for Development of Advanced Computing(C-DAC), Mohali, India

*ABSTRACT*

*In today's VLSI system design, power consumption is gaining more attention as compared to performance and area. This is due to battery life in portable devices and operating frequency of the design. Power consumption mainly consists of static power, dynamic power, leakage power and short circuit power. Dynamic power is dominant among all which depends on many factors viz. power supply, load capacitance and frequency. Switching activity also affects dynamic power consumption of bus which is determined by calculating the number of bit transitions on bus. The purpose of this paper is to design a bit transition counter which can be used to calculate the switching activity of the circuit nodes. The novel feature is that it can be inserted at any node of the circuit, thus helpful for calculating power consumption of bus.*

## 1. INTRODUCTION

In recent years, with the development of VLSI technology, Integrated Circuits (IC) come into existence. More and more sophisticated systems are being implemented on Integrated Chips. Higher levels of integration and shrinking line widths have led to a generation of devices which are more sensitive to power dissipation and reliability problems [1]. These systems consume considerable amount of energy and power. Power consumption is becoming a critical concern in designing and testing field as compared to performance and area. This is due to two main factors viz. battery life in portable devices and operating frequency. The other factor is the heat dissipated due to large currents within the circuit, which must be removed by proper cooling techniques.

Power consumption mainly consists of static power and dynamic power. Static power is due to leakage current in the circuit when circuit is in static state i.e. not working, there is still a leakage current flowing in the Complementary Metal Oxide Semiconductor (CMOS) operating in sub threshold region. The leakage current is given by:

$$i_l = i_s(e^Q - 1) \qquad (1)$$

where $Q=q^{v/kT}$, $i_s$ is reverse saturation current, V is diode voltage, q is electronic charge, k is Boltzman constant and T is temperature. The product of this leakage current with supply voltage gives the value of Static Power dissipation for circuit. If there are n numbers of devices in the circuit then sum of leakage current of all the devices is taken into consideration to give the average static power dissipation in the circuit as:

$$P_s = (\text{leakage current} * \text{supply voltage}) \qquad (2)$$

Dynamic power is due to switching transient current and charging & discharging of load capacitance [2]. It depends on many factors like load capacitance, supply voltage and frequency of operation. Average dynamic power dissipation in circuit is given by:

$$P_d = C_L * V_{dd} * f \qquad (3)$$

The one important factor which also affects the dynamic power consumption is switching activity of nodes. Transition Density [1] which is average number of transitions per second at a circuit node is a measure of switching activity. Switching activity is defined as the rate of switching of the circuit from '0' to '1' and '1' to '0' while operating in the circuit. This switching is taken into consideration whenever bus power consumption is calculated. Generally switching activity occurs at node and difficult to evaluate, estimation model are given in [11]-[12]. Switching activity can be determined by calculating the number of bit transitions takes place during actual operation of logic in circuit. Gray Code [13] is the simplest technique used to reduce the number of transitions while various other techniques are also used like Bus Invert [14]. Switching activity also called as node transition rate can be slower than the clock rate. Hence a node transition factor or switching activity factor ($\alpha_T$) must be introduced in (3), which is given by:

$$\alpha_T = \text{No. of bit transitions} / \text{Total No. of bits} \qquad (4)$$

The maximum value of $\alpha_T$ is 1, as the number of transitions cannot exceeds the total number of bits. For example in 8 bit data if the bit transitions of the new data are 2 from the previous data than $\alpha_T = 2/8 = 0.25$. By introducing $\alpha_T$ factor in the (3), the average dynamic power dissipation on bus is given by:

$$P_d = \alpha_T * C_L * V_{dd} * f \qquad (5)$$

It is clear from above equation that by minimizing the switching activity, the average dynamic power consumption on bus can be reduced as it is directly proportional to $\alpha_T$. For some tests it is possible to bound the switching activity such that it would not exceed that possible during functional operation [9]. A lot of research work is going on to reduce dynamic power consumption on bus by reducing the switching activity [3]-[4]. Low power address generators are proposed in [5][8], which mainly focus on reducing switching activity. Also research work is going on reducing power through charging and discharging of node capacitance [6].

The paper is organized as follows. Section 2: outlines the need and designing part. Section 3: describes the implementation part. Section 4: shows the results and Section 5: draws the conclusion.

## 2. DESIGNING OF BIT TRANSITION COUNTER

Before designing Bit Transition Counter (BTC), first we decide the specifications of BTC or with what requirements it should be designed. As discussed earlier the switching activity plays

Circuits and Systems: An International Journal (CSIJ), Vol. 1, No. 1, January 2014

important role in dynamic power consumption. Hence it is important to determine $_T$ while going for low power design. Also switching takes place not only at one node of the circuit but at all nodes, make difficult to calculate all switching simultaneously. Switching activity at the nodes is responsible for bus power consumption. Dynamic Power consumption is not limited to designing part in VLSI, but also in testing part it is of great concern. A considerable amount of power is consumed while testing digital circuits. Hence low power testing is needed. The different solutions to reduce power consumption are while testing a digital circuit low switching test pattern generators are preferred. The low transition activity is preferred in Neural Networks [10]. Also for memory testing low switching address generators are preferred.

Considering all above points in considerations the BTC is designed which is capable of counting the number of bit transitions in the successive data. For example if data changes from "00111100" to "11111101" the bit transitions are 3 as transitions takes place at data(7), data(6) and data(0). The BTC counter should be capable of counting number of transitions at every transition between successive data.

The block diagram is shown in Fig. 1. As shown the input is clock, reset, datain of 16-bit and output as dataout of 16bit. The datain is the input data at which the transition occurs, the output data is same as input data as it should not be changed. The datain and dataout can be changed to any number of bit widths as per requirements. The other outputs are one_transition and total_transition. The one_transition count the number of bit transitions between two successive data and then resets to zero value to indicate the next bit transition value between next two successive data. The total_transition counts the number of total transitions takes place while the circuit is operating. The output of total_transition can be obtained by adding the values of one_transition at every clock cycle, thus obtaining the total number of transitions during the circuit operation.

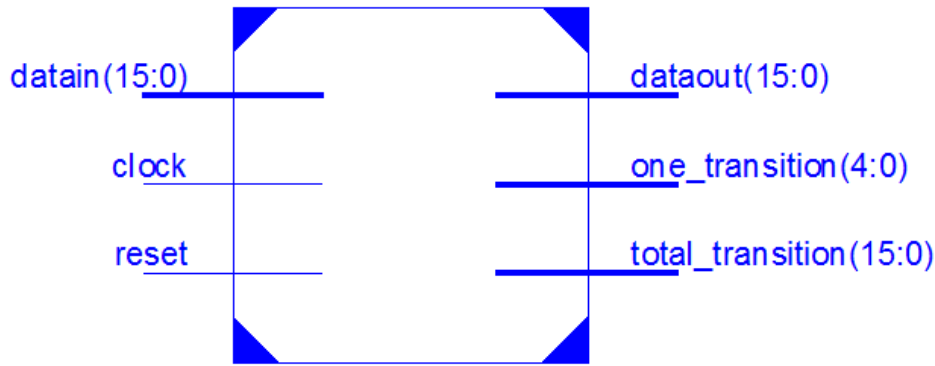

Figure1. Block diagram of Bit Transition Counter

## 3. IMPLEMENTATION OF BIT TRANSITION COUNTER

As described the BTC is designed so that it is capable of counting the number of bit transition between two successive data without affecting the actual data. So, it can be implemented at any section of the circuit or at any node of the circuit. As shown in Fig. 2 the bit transition counter is implemented on the basic digital circuit while testing. Basic digital testing circuit [7] consist of Test Pattern Generator (TPG), Circuit under Test (CUT) and Output Response Analyzer (ORA). The bit transition counter can be implemented at any section, but we have implemented it after the test pattern generator to calculate the number of transitions made by the test pattern generator.

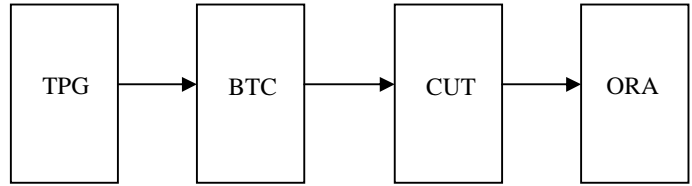

Figure 2. Implementation of Bit transition counter

As shown in Fig. 2, we have implemented BTC on different types of test pattern generators which consist of Internal Linear Feedback Shift Register (LFSR), External LFSR, Cellular Automata (CA)-90 and Cellular Automata-150. The bit transition counter calculates the number of transitions takes place while applying the test patterns by these TPG's. By using these transitions the switching activity is calculated which is shown in the results section. Also the BTC is implemented on the address generators i.e. Binary Counter and Gray Counter and the same results are shown in next section.

## 4. EXPERIMENTAL RESULTS

The BTC is designed in VHDL and implemented on Cadence Design Tool Suite. The code is synthesized in rc tool and the various reports are generated such as power, area and Fanout & Net analysis as highlighted in Table 1, 2 and 3.

Table 1. Power Analysis

| Power type | | Power dissipation (uw) |
|---|---|---|
| Static | | 1.3 |
| Dynamic | Internal | 8.3 |
| | Net | 1.3 |
| Total | | 10.9 |

Table 2. Area Analysis

| Leaf type | Cells | Cell Area | Area (%) |
|---|---|---|---|
| Combinational | 32 | 90.317 | 24.2 |
| Sequential | 16 | 282.240 | 75.8 |
| Total | 48 | 372.557 | 100 |

Table 3. Fanout and Net Analysis

| Fanout & net ratio type | Value |
|---|---|
| Maximum Fanout | 16 (clock) |
| Minimum Fanout | 1 (datain[0]) |
| Average Fanout | 2.0 |
| Terms to net ratio | 2.1 |
| Terms to instance ratio | 3.0 |

The simulation of the VHDL code is performed in ncsim tool and the waveforms are shown in Fig. 3. As shown the datain is the input data from any source and the dataout is the output data which is same as the input data, but having value after one clock cycle. When the reset input is high the outputs one_transition and total_transition gets reset to 0 value. When reset input goes low then the one transition and total transition counts the value of number of transitions as the input changes. As in our case the input data is 0000 (hex) or "0000000000000000" initially and the next input is 0303 (hex) or "0000001100000011", hence the number of bit transitions are 4 which is same as the one_transition output value i.e. 04 (hex) and same as total_transition 04 (hex), the one_transition output get reset to 0. At next transition the input data is 0F03 (hex) or "0000111100000011", hence the number of bit transitions is 2 which is shown by the one_transition output and total_transition as 6 (4+2). The same procedure repeats till the circuit operates for clock cycles. The total_transition output will give the total number of transitions takes place in the circuit.

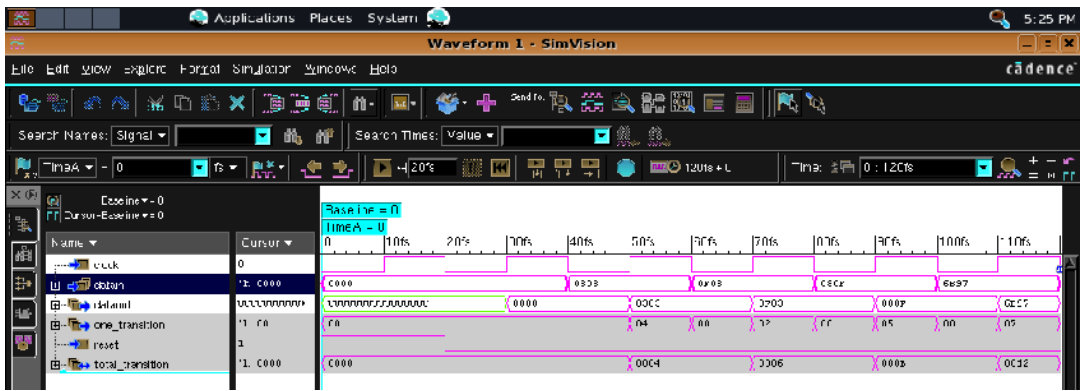

Figure 3. Simulation Results of Bit Transition Counter

As discussed earlier the BTC is implemented on the address generator in memory testing i.e. the binary counter and gray counter. The number of transitions and switching activity for 4-bit and 8-bit binary counter and gray counter is shown in Table 4. It is clear that by using gray counter the switching activity gets reduced by almost 40-50%. While performing simulations it is observed that number of transitions with successive clock cycles gets reduced in gray counter, hence with more clock cycles the switching activity gets reduces by more factor.

Table 4. Switching activity of Binary and Gray counter

| Module | No. of transitions | Switching Activity |
|---|---|---|
| Binary Counter (4-bit) | 26 | 0.43 |
| Gray Counter (4-bit) | 15 | 0.25 |
| Binary Counter (8-bit) | 502 | 0.246 |
| Gray Counter (8-bit) | 255 | 0.125 |

Also it is implemented on test pattern generators such as Internal LFSR, External LFSR, CA-90 and CA-150. The number of transitions and switching activity for the same is shown in Table 5. The initial seed pattern is taken as "1011001010110110" in all cases and then patterns are generated by giving clock, at every clock new pattern is generated according to the logic defined and the transitions between the bits takes place. Bits transitions are calculated by the bit transition counter, with the help of it the switching activity is calculated. The results are taken for 8, 16 and 32 clock cycles as highlighted in Table 5. The results show that internal LFSR accounts for highest switching activity among 4.

Table 5. Switching Activity of Pattern Generators

| Module | No. of transitions | | | Switching Activity | | |
|---|---|---|---|---|---|---|
| | *Clock Cycles* | | | | | |
| | *8* | *16* | *32* | *8* | *16* | *32* |
| Internal LFSR | 66 | 114 | 236 | 0.51 | 0.44 | 0.46 |
| External LFSR | 88 | 163 | 266 | 0.68 | 0.63 | 0.51 |
| CA-90 | 66 | 138 | 276 | 0.51 | 0.53 | 0.53 |
| CA150 | 67 | 135 | 259 | 0.52 | 0.52 | 0.50 |

## 5. CONCLUSION

In this paper, we proposed the designing of BTC for determining the switching activity which is responsible for dynamic power consumption of bus in the circuit. BTC is successfully implemented on address generators of memory testing and test pattern generators for digital circuit testing. Our approach can be implemented at any node of the circuit which is also helpful in determining the dynamic power consumption and is directly related to node transition activity.

## Authors

**Balwinder Singh** has obtained his Bachelor of Technology degree from National Institute of Technology, Jalandhar and Master of Technology degree from University Centre for Inst. & Microelectronics (UCIM), Punjab University, Chandigah in 2002 and 2004 respectively. He is currently serving as Senior Engineer in Centre for Development of Advanced Computing (CDAC), Mohali and is a part of the teaching faculty and also pursuing Phd from GNDU Amritsar. He has 6+ years of teaching experience to both undergraduate and postgraduate students. Singh has published three books and many papers in the International & National Journal and Conferences. His current interest includes Genetic algorithms, Low Power techniques, VLSI Design & Testing, and System on Chip.

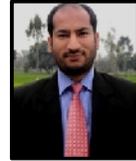

**Amandeep Singh** received the B.Tech. (Electronics and Communication Engineering) degree from the Beant College of Engineering and Technology, Gurdaspur affiliated to Punjab Technical University, Jalandhar in 2010, and M.Tech. (VLSI design) degree from Centre for Development of Advanced Computing (CDAC), Mohali in 2012. Presently he is doing Phd at National Institute of Technology, Jalandhar. His area of interest is VLSI Design and Testing, Device Modelling, FPGA based Design.

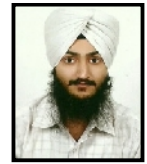